# Ethical Issues Regarding the Use of AI Profiling Services for Recruiting: The Japanese Rikunabi Data Scandal

Fumiko Kudo [1], Hiromi Arai [†2,3], Arisa Ema [‡1,2]

[1] Institute for Future Initiatives, The University of Tokyo, Japan
[2] Center for Advanced Intelligence Project, RIKEN, Japan
[3] JST PRESTO

## Abstract

The ethical, legal, and social challenges involved in the use of profiling services for recruitment are the focus of many previous studies; however, the processes vary depending on the social system and cultural practices. In August 2019, a scandal occurred in Japan in which a recruitment management company was found to have breached users' and students' trust by selling their data to clients. By sharing the Japanese recruitment context and associated laws, this article contributes to our understanding of the ethical issues involved in artificial-intelligence (AI) profiling and in handling sensitive personal information.

## Introduction

There is growing interest in Human Resource (HR) technology that uses artificial intelligence (AI) for recruitment and personnel affairs, including in Japanese companies. In a survey (n = 400) of HR professionals conducted in 2019 at companies with annual sales of 50 billion yen or more, or with 1000 or more employees, more than 60% of the participants responded that they had positive expectations from introducing such technology, including those who had already introduced it[1]. Among companies that have already introduced or are planning to introduce such a system in the future, "recruitment" was the area in which they most desired to apply this technology. Specifically, "entry sheet acceptance, content confirmation, and screening" were the most popular processes for application.

Despite the positive expectations attached to using AI technology in the recruiting process, there are concerns about profiling. Specifically, among the ethics of algorithms, profiling by algorithms is considered one source of discrimination (Mittelstadt et al., 2016). For example, Amazon's recruitment AI, which was developed based on existing hiring data, was abandoned because of its discrimination against women[2]. It is suggested that predictive hiring tools can reflect institutional and systematic biases; therefore, digital sourcing platforms should recognize their influence on the hiring process (Bogen and Rieke, 2018). However, the hiring process depends on social systems and customs, and the issues may vary according to country.

## The Recruit DMP Follow Incident

In August 2019, a scandal occurred in Japan when a recruitment management company was found to have betrayed users' and student's trust by selling their data to client companies. The service is named "Recruit Data Management Platform (DMP) follow". This section will provide an overview of the social, systematic, and legal background by carefully extracting the issues related to this incident.

### Social Context: The Need for Profiling Services

Recruit Career Co., Ltd. (hereinafter referred to as "Recruit Career"), is a major recruitment management company in Japan, and is part of the Recruit Holdings Co. group. It operates one of the largest job placement websites, "Rikunabi," which matches employers (companies) with job seekers (students) in Japan.

Japan has a custom called "simultaneous recruitment of new graduates," in which students in the same grade all seek jobs at the same time. Accordingly, Rikunabi provides website data by year (e.g., "Rikunabi X"; X is the year of graduation, such as 2019), and students register on their graduation-year sites to use the Rikunabi service. For "Rikunabi 2020," the number of listed companies is 31,564 and that of students is approximately 800,000.

With the simultaneous recruitment of new graduates, some students may receive informal job offers from several companies. Consequently, the issue of "declining job offers" arises. It is difficult for companies to find alternative

---

[†] hiromi.arai@riken.jp, [‡] ema@ifi.u-tokyo.lac.jp
This paper was written in November, 2019.

[1] prtimes.jp/main/html/rd/p/000000021.000026061.html (Accessed October 26, 2019; note: all the websites in this article were accessed on the same day).

[2] www.reuters.com/article/us-amazon-com-jobs-automation-insight/amazon-scraps-secret-ai-recruiting-tool-that-showed-bias-against-women-idUSKCN1MK08G

employees if students decline their job offer because most candidates already have job offers. This trend is spurred by the "employee-dominated market," caused by a decrease in the working population in Japan and the active job market. According to the Ministry of Health, Labour and Welfare (MHLW), the employment rate of university students graduating in March 2019 was 97.6%, the 2nd highest since the survey started in 1997[3]. As a result, there is a growing need among companies to understand the applicants' probability of accepting to work in the company. According to media reports, a personnel management officer said, "It is helpful to know who will decline an offer 50% or 5% of the time during the recruiting process. With that valuable data, recruiters can develop a data-driven recruitment strategy. It is efficient to focus on recruiting students who have a 5% chance of declining rather than 50% if their evaluation is equivalent.[4]"

**System and Data Used: Profiling Service Specifications**

Recruit Career inaugurated a profiling service called "Rikunabi DMP follow" in March 2018. This service collects and analyzes demographic information and cookies of job seekers (students) collected through the matching service and calculates the probability score of students declining informal job offers (hereinafter referred to as the score) for a specific company.

The score was calculated for 74,878 users and sold to 34 companies (of the 38 companies that had signed up for this feature). We will explain the data and system used in this process based on the external explanation document released by Recruit Career[5].

The entities and shared data are as follows. The three entities involved are Recruit Career, clients of Rikunabi DMP follow, and Rikunabi X users (figure 1).

Recruit Career and a client company shared users' data, including each user's ID and affiliation (university, faculty, and department), information taken by the client company, and user's cookies. Recruit Career performed cookie synchronizations (cookie syncs); however, individuals were not identifiable by the shared IDs when they launched the service in March 2018. However, starting March 2019, Recruit Career and the client companies also shared the users' identifiers and Recruit Career performed cookie syncs with identified information.

The algorithms used for profiling are uniquely designed for each company. They are derived from a comparison of the users who accepted the offer and those who declined the offer the previous year. The specific algorithms are not published.

The value of the delivered score was in the range of 0.0 to 1.0, and the possibility of leaving the selection or declining the offer was calculated for individual users. When the score value could not be calculated, or was blank or "N/A," no score was given. A sample of the scoring is shown in table 1[6]. Note that the value calculated as the score is not a percentage. For example, "Score 0.4" does not mean "a 40% chance of declining." The scores were then divided into five categories using star (★), according to the needs of the client companies.

| Student ID | Score | Probability of declining |
|---|---|---|
| 10001 | 0.4 | ★★ |
| 10002 | 0.53 | ★★★ |
| 10003 | 0.61 | ★★★ |
| 10004 | 0.23 | ★★ |
| 10005 | 0.1 | ★ |

Table 1 A sample of the score provided by Recruit Career.

Recruit Career insists that the score is not intended to be used for selection and acceptance decision making, stating, "We ask companies to promise that they will not use the data to judge acceptance or rejection." The service is to be used only for promoting communication between users and companies, such as to follow up after a job offer.

**Legal Defect: Inadequate Privacy Policy**

The consent of the user is required prior to sharing any information about them, in accordance with the Act on the Protection of Personal Information (hereinafter called "APPI"), as this profiling service provides users' personal information to third parties (client companies). However, the privacy policy was inadequate and 7,983 users did not give valid consent.

Recruit Career tried to change its privacy policy to one that referred to "Rikunabi DMP follow" in March 2019, but the change was not reflected on some screens (Mislabeling)[7]. Furthermore, it was not designed to obtain the appropriate consent from all users who are subject to the score (insufficient consideration of agreement acquisition flow). As a result, a total of 7983 users' information ("those who have not used the job-seeking function of the site since March 2019" and "those whose scores were ob-

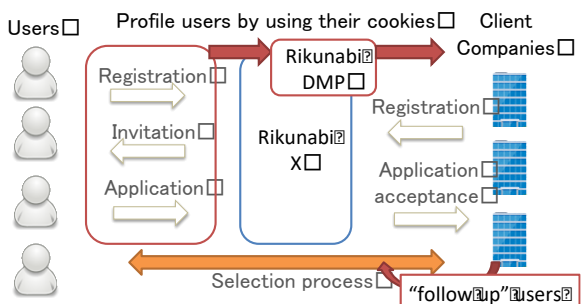

Fig1. The entities and shared data

---

[3] www.mhlw.go.jp/stf/houdou/0000205940_00002.html
[4] business.nikkei.com/atcl/gen/19/00002/080200592/
[5] www.recruitcareer.co.jp/news/pressrelease/2019/190826-01/

[6] www.recruitcareer.co.jp/news/pressrelease/2019/190826-01/
[7] "Rikunabi X" has a site configuration in which users (student) agree to the privacy policy on multiple screens.

tained after March 2019 among applicants to companies that have introduced the Rikunabi DMP Follow") was provided to the company without appropriate consent.

The privacy policy displayed on the membership registration screen is as follows:

> Cookie information obtained by this service or sites affiliated with our company (Recruit Career) is analyzed and collected, and that information may be provided to clients for optimal information provision and recruitment activities for users (It is not used for selection). If the user discloses personal information, our company may distribute and display advertisements and contents, support recruitment, and provide the service by using users' action history from before the information disclosure.

The privacy policy that was originally scheduled to be published was as follows:

> When a user logs in and uses the service, after identifying the individual, the personal information registered by the user with the service and activity histories (including the activity history from before the login) obtained from the service or sites affiliated with our company (Recruit Career) using cookies may be analyzed and collected, and used for the following purposes: (1) Optimal provision of information to users, such as distribution and display of advertisements and content, and (2) Provision of information to companies using the system to assist recruitment activities (it will not be used for the selection.)

## Incident: Press Reporting and Guidance

This section outlines how the incident garnered attention by focusing on three entities: Recruit Career, the press, and the government.

**Discovery: Investigation by Personal Information Protection Commission and the Media Scoop**
The profiling service gained public attention and became a social issue because of a covert administrative investigation and press coverage. The time series is as follows (figure 2).

On July 9, 2019, the Personal Information Protection Commission (hereinafter referred to as "PPC") held a closed hearing with Recruit Career. Following the hearing, Recruit Career conducted an internal investigation and decided to suspend "Rikunabi DMP follow" on July 31.

One influential Japanese newspaper, Nihon Keizai Shimbun, noticed the move and published an article on the evening of August 1, entitled "[evening scoop] Rikunabi provides 'information on the probability of students declining informal job offers' without explanation[8]." This scoop led to a public outcry. Later that night, Recruit Career issued a press release entitled "Regarding some reports on our company services[9]." At that time, there was no clear recognition of breaches of the APPI and the company only mentioned that "this service will be temporarily suspended until we have considered more detailed explanations," suggesting the possibility of restarting the profiling service.

**Exposure: Survey by the Tokyo Labour Bureau and Announcement of Service Abolition**
Following Nihon Keizai Shimbun, other media outlets also reported the incident. Critical comments were also posted on social media.

On August 2, the Tokyo Labour Bureau (hereinafter referred to as "TLB"), a local branch of the Ministry of Health, Labour and Welfare (MHLW), conducted a confidential investigation on Recruit Career[10]. The same day, the Minister of Education, Culture, Sports, Science and Technology (MEXT), Masahiko Shibayama (at that time) said at a press conference "It may have been unexpected from the student's point of view that information that had a very large effect on job hunting was provided to companies without the students' knowledge."[11]

On August 5, Recruit Career released a document "Regarding the inadequate consent obtained through the privacy policy in the case of 7983 students at 'Rikunabi DMP follow' and the abolition of the service[12]." The company apologized for the flaws in its privacy policy as found by an internal investigation and declared service abolition. The company positioned the incident as a betrayal of users' (students') trust, rather than as a breach of the APPI. This betrayal can be seen from the company's statement that noting "our company's lack of awareness of the feelings of the students has led us to recognize that this is a fundamental issue." It is inferred that such changes in perceptions were influenced by surveys conducted by the TLB, statements made by the Minister of MEXT, and public opinion through social media.

**Aftermath: Spillover to Client Companies**
In addition to Recruit Career, criticism was also directed at client companies that purchased the data. On August 8, Minister of MHLW, Takumi Nemoto (at that time) said that 38 client companies that purchased the data were also subject to investigation, and said, "strict guidance should be given if there is a violation of personal information[13]."

Regarding the social network response, many of the replies to tweets from the media accounts directly criticized Recruit Career. There were many requests to disclose the names of the client companies that had purchased the data, before this information could become publicly available.

---

[8] www.nikkei.com/article/DGXMZO48076190R00C19A8MM8000/
[9] www.recruitcareer.co.jp/news/pressrelease/2019/190801-02/
[10] www.nikkei.com/article/DGXMZO48289790W9A800C1916M00/
[11] www.jiji.com/jc/article?k=2019080200671&g=soc
[12] www.recruitcareer.co.jp/news/pressrelease/2019/190805-01/
[13] www.nikkei.com/article/DGXMZO48354670Y9A800C1EAF000/, English article is available from Japan Times as well (www.japantimes.co.jp/news/2019/09/06/business/corporate-business/ministry-says-recruit-career-acted-illegally-rikunabi-data-scandal-demands-corrective-steps/#.XbP8C-j7Sbh)

On August 9, the name of the company that purchased the scores was first revealed, and it was Honda, a major auto company[14]. After that, famous companies such as Toyota, YKK, Mitsubishi Electric, Aflac Life Insurance, and Resona Holdings were also reported to have purchased the scores. The most popular use as stated by the companies of the scores was to follow up with applicants and candidates (e.g., Aflac Life Insurance, Resona HD, YKK, Kyocera and other 9 companies). Other purposes were technical verification for future use of artificial intelligence in recruitment activities (Daiwa Institute of Research Group), and behavioral analysis of applicants who declined offers (Mitsubishi Electric). The companies explained how they dealt with the data and the applicants through press releases.

**Termination: Recommendations and Guidance Led to Apology via a Press Conference**

On August 26, the PPC declared actions of "recommendation" (APPI, Article 42, Paragraph (1)) and "guidance" (APPI, Article 41)[15]. It urged the company to take corrective measures and use personal information appropriately.

In response to the recommendations and guidance, Recruit Career held an apology press conference on the same day. Daizo Kobayashi, the president of Recruit Career, admitted there was a lack of consideration of the students and a failure of internal governance, and apologized for any inconvenience caused to students, universities, and corporate personnel[16]. Recruit Career also released a document entitled "Recommendations to our company concerning 'Rikunabi DMP follow'[17]." In the press conference and release, the facts and background of this incident were described as well as measures the company had taken to strengthen governance.

On September 6, the TLB also declared guidance to Recruit Career, based on the Employment Security Act. In response, Recruit Career issued a press release "Regarding guidance by the TLB for our company concerning 'Rikunabi DMP follow'[18]". The release said the TLB and the PPC investigations are to continue.

## Discussion

The scandal contains various perspectives, so we divided it into two legal aspects and ethical and social implications.

**Personal Information Protection**

As mentioned above, the PPC declared "recommendation" and "guidance." The "facts underlying the recommendations" are (1) flaws in security control actions at the time of service design, (2) flaws in security control actions at the time of changes in the privacy policy, and (3) lack of consent regarding third-party provision[19].

**Lack of Security Control during Design:** In the recommendation, the PPC declared that Recruit Career has breached the APPI, Article 20, which means that it failed to take the necessary and appropriate actions for ensuring the security of personal information including preventing the leakage and loss or damage of such information, in the context of the services of 'Rikunabi DMP follow'.

However, Recruit Career took this profiling service to the R&D stage; therefore, it was considered different from ordinary service development. In addition, there was a lack of collaboration between the legal department of Recruit Career Co. Ltd, a subsidiary company, and the legal department of Recruit Holdings Co. Ltd., the parent company. As a result, the service passed through the approval process without adequate verification, primar-ily led by Recruit Career Co., Ltd., which is a more business-oriented interest.

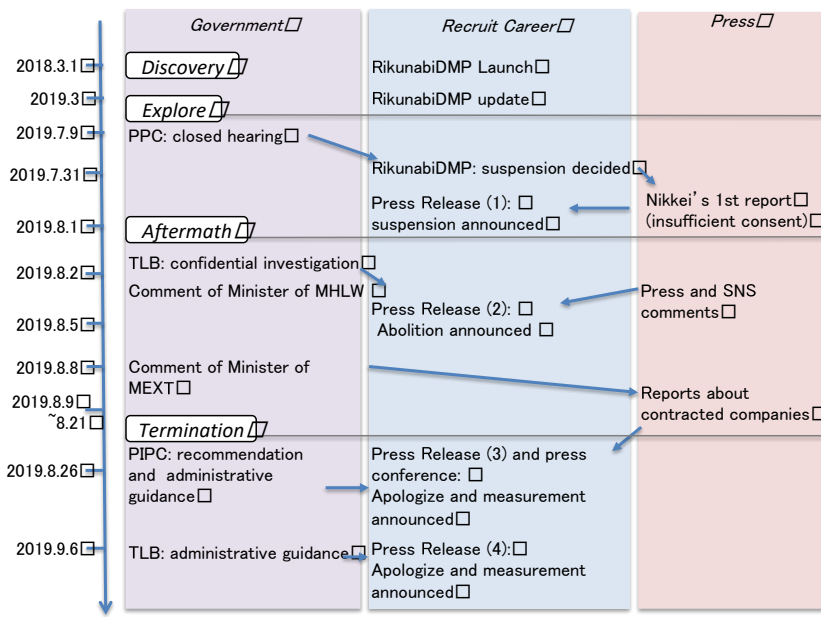

Fig.2 The time series of the incident

---

[14] www.nikkei.com/article/DGXMZO48453300Z00C19A8EA5000/?n_cid=DSREA001. English article is available from the Japan Times (www.japantimes.co.jp/news/2019/08/14/business/corporate-business/toyota-honda-bought-data-based-job-hunters-browsing-activity/#.XbP7_Oj7Sbh)
[15] PPC said in its first instructions for corrective action since its establishment in 2016
[16] www.itmedia.co.jp/news/articles/1908/27/news062.html. English article is available from the Japan Times (www.japantimes.co.jp/news/2019/08/27/business/corporate-business/recruit-career-job-data-scandal/#.XbP8A-j7Sbh)
[17] www.recruitcareer.co.jp/news/pressrelease/2019/190826-01/
[18] www.recruitcareer.co.jp/news/pressrelease/2019/190906-01/
[19] www.recruitcareer.co.jp/news/pressrelease/2019/190826-01/

Some experts are concerned about the lack of personal information segregation. Recruit Career should have distinctively managed each piece of personal information obtained by the client companies and should not have merged personal information obtained from one client (e.g. Toyota) with personal information obtained from another client (e.g. YKK).

**Lack of Security Control Measures when Changing Privacy Policy:** The PPC also declared in its recommendation that the company had violated the APPI, Article 20, for inappropriate information management, such as the revision of the privacy policy. When the privacy policy was revised in March 2019, because of inadequate procedures, it did not obtain students' consent to the provision of personal information to third parties. Thus, Recruit Career did not have a structural system to prevent, detect, and correct this defect and it failed to find it until the PPC inspection. In other words, 'Rikunabi' is an annual service, and there is usually no change after the release of such a service. As for the irregular revision, the necessary workflow and steps to be taken were not prepared, resulting in the omission of the privacy policy on some screens. Recruit Career admits that its privacy controls are flawed.

**Lack of Consent to Provide Information to a Third Party:** The PPC also mentioned that Recruit Career has breached the APPI, Article 23, Paragraph (1). "When providing personal information to a third party, it is necessary to obtain consent. However, personal information of 7,983 students was provided to the third-party company without their consent[20]." In light of this, the PPC told Recruit Career to clearly explain to Rikunabi users how it offers their personal information to third parties. Specifically, reasonable and appropriate content, such that it is can enable a person to make an informed decision on consent, shall be provided.

The PPC also mentioned that "The privacy policy failed to clearly explain the process whereby the company provides personal information to a third party in the 'Rikunabi DMP follow." However, it is unclear whether the "privacy policy" in this recommendation refers to the one that appears on the screen, or the one that was originally planned to appear. If the latter is included, the problem is the explanation about the cookie sync is insufficient and the score content is not fully explained to the user.

**Employment Security Act**
The case also raises labor law issues. Guidance by the TLB determined that Recruit Career's sharing of personal information with client companies violated the Employment Security Law (hereinafter referred to as "ESL"). TLB guided that Recruit Career should be confirmed not to have violated the ESL and related regulations. It also should take corrective steps and preventive measures, for example, to improve business operations and systems. In addition, TLB instructed Recruit Career to respond sincerely to inquiries from students whose information has been shared in the 'Rikunabi DMP follow' and to take appropriate measures, such as providing a careful explanation of the information and its provision.

The guidance wording is ambiguous, so it needs interpretation. The ESL and related regulations override some parts of the APPI regulation and employment placement business providers are obligated to properly manage the personal information of job seekers (ESL, Article 5-4, Paragraph (2)). According to its guidelines, "employment placement business providers shall collect personal information by lawful and fair means." In addition, as proper personal information management (Guideline 4-2), in cases where a business provider has come to know secret personal information of a person who intends to become an employee, strict management shall be exercised so that said personal information will not be disclosed to others without justifiable grounds.

In this context, "Personal Information" refers to any information that can identify an individual, of which "secret" refers to "a fact that is not generally known (the requirement of protection) and that is objectively found to be of reasonable benefit to the person in relation to what is not known to others (the requirement of nonpublic)." Specifically, "Your permanent address, place of origin, party of support or affiliation, history of political movements, amount of borrowings, and the fact that you are a guarantor may be confidential." The TLB has not brought out whether user's cookies and the score fall under the category of "secret". However, because there may be a conflict of interest between the students and the companies, it is likely that there are reasonable benefits for the students (regarding the requirement of protection). Moreover, students are unlikely to disclose information to the public when they are hesitating to decline an offer. Therefore, the score may be included in the category of "secret" (regarding the requirement of nonpublic) and deserve a high level of protection. However, there is another view that the score is not the personal information itself, but evaluations. This point will be mentioned later.

TLB guidance also emphasized that "the main purpose of the employment placement business" is to "make an intercession between job offerors and job seekers after receiving applications for job openings and job applications" (ESA, Article 4, Paragraph 1). The term "intercession" means "to take care of the job offerors and the job seeker as a third party to facilitate the establishment of the employment relationship". Consequently, support is likely to be fair and neutral, utilizing an intercessor between job seekers and employers. From the guidance, it can be understood that Recruit Career should have paid attention not only to the benefit of client companies but also to the students who sought employment.

---
[20] www.recruitcareer.co.jp/news/pressrelease/2019/190826-01/

**Ethical Concerns on Power Structure**

As illustrated in figure 1, the structure of the incident occurred among trilateral relationships of data subject (users/students), analyzer (Recruit Career) and recipient of the profiled data (client companies). The incident illustrates that Recruit Career designed the profiling service without considering the data subjects' benefits. Users and students are in a vulnerable position, and therefore, the analyzer and the client companies should recognize the power structure. Contrary to the previous studies on profiling services that focuses on data and algorithmic biases and discrimination against people of race and gender in hiring (Kim, 2017), this incident raises compliance issues of the platform company and the recipient of the profiled data companies.

However, even though the legal issues mentioned above have been cleared, ethical issues remain. User trust in Recruit Career has been betrayed, resulting in approximately 40% of students who are currently and will be seeking jobs expressing that they will reduce the use of Rikunabi X services, according to the questionnaire survey conducted by Nihon Keizai Simbun after the incident[21]. Also, some universities announced they will not invite Recruit Careers representatives as lecturers at the university[22] or introduce students to the services[23].

**Social implications on HR Tech**

In relation to this incident, media articles explain that "Rikunabi DMP Follow" is a service offering student probability scores of declining informal job offers omitting the word "artificial intelligence". Although it is not possible to determine what kind of algorithms were used based on the publicly available materials, the service was probably not as sophisticated as using machine learning.

However, some media outlets write that "artificial intelligence predicts the probability of declining an offer from various components of personal data." In this respect, this is the first case in which "artificial intelligence" has become a social problem when it is used in personnel affairs and recruitment in Japan. Due to this publicity, some engineers are concerned that the idea that artificial intelligence used in recruitment might be linked to the negative image in the public.

However, the main issue with this incident was the legal defect on personal information usage, the company handling the data lacked consideration for personal information.

On the contrary, HR Tech using machine learning is already being used. Recruit Career Group noted in a post in September 2018 that they were developing and deploying machine learning for its own recruitment on students, and that it expects to sell HR Tech services to others in the future[24].

---

[21] headlines.yahoo.co.jp/article?a=20191011-00000002-nikkeisty-bus_all
[22] www.tsukuba.ac.jp/public/newspaper/pdf-pr/351.pdf
[23] www.nikkei.com/article/DGXMZO49111660Y9A820C1TJ1000/
[24] www.itmedia.co.jp/business/articles/1809/20/news074_3.html

# Conclusion

The probability score of students declining informal job offers and analysis, evaluation and predictions are based on personal information, and may not be the personal information itself. On a practical level, too much regulation or restriction on analysis can make meaningful profiling difficult, creating challenges for developers or companies to accept. However, the higher the analysis accuracy, the more likely it is to be accepted as true, resulting in a higher risk of violating the legal interest on the data subject.

Regarding the balance between the analysis accuracy and the protection of personal data, no concrete guidelines have been presented at this stage, therefore ethical and social discussion becomes important. In Japan, a research group released "Draft Recommendations on Profiling Governance" (the Personal Data + Alpha Research Group, 2019). It suggests (1) the companies establish a compliance system for profiling, (2) industry associations support voluntary profiling efforts, and (3) the government support voluntary initiatives led by the private sector. It also created a "checklist for self-regulation" at each of the four planning stages, data acquisition, processing and operation. The checklist pointed out the potential conflicts of interest among data subjects, analyzes and recipient of the profiled data. It also emphasizes ensuring an adequate explanation of the purpose of profiling to the data subject before rendering services.

In addition, these guidelines need to be in line with preexisting general guidelines (Jobin et al, 2019). In October 2019, Recruit Works Institute, a research institute of Recruit Holdings Co., Ltd., published a special issue on its bulletin featuring draft AI principles on AI (Irikura, 2019). It consists of four fundamental principles (respect for the rights of individuals, sensitivity to inputs (personal information), accountability for the output, and professional ethics and literacy as a personnel management officer) and 15 specific principles. To create the draft, not only were Japanese principles referenced, such as the Social Principles of Human-Centric AI by the Japanese Cabinet Secretariat and AI ethical guidelines by the Japanese Society for AI, but international principles, such as OECD principles, on AI and GDPR are referenced as well.

For consensus building among stakeholders, guidelines on human resource technologies and, more specifically, profiling guidelines have become essential. These guidelines need to be in line with the global AI governance principles as well as national or sectoral context and customs. Although this incident occurred in Japan, we hope that the case and the discussion contribute to the governance and ethics of AI profiling and handling sensitive personal information.